\documentclass{article}

\usepackage{graphicx}
\usepackage{caption}
\usepackage{subcaption}
\usepackage{booktabs}
\usepackage{adjustbox} 
\usepackage{rotating}
\usepackage{amsmath, amssymb}
\usepackage{microtype} 
\usepackage{float}
\usepackage {natbib}
\usepackage[colorlinks=true,allcolors=blue]{hyperref}
\usepackage[T1]{fontenc}
\usepackage{lmodern}
\newcommand{\blind}{0}

\begin{document}

\def\spacingset#1{\renewcommand{\baselinestretch}%
{#1}\small\normalsize} \spacingset{1}


\if0\blind
{
 \title{\bf Bayesian Joint Estimation of the Hurst Parameter and Volatility with Applications to Fractional Option Pricing}
  \author{
    Hana H. Sagor$^{1}$, Edward L. Boone$^{1}$,  Ryad A. Ghanam$^{2}$ \\
    \\
    $^{1}$Department of Statistical Sciences and Operations Research, \\
    Virginia Commonwealth University, Richmond, Virginia, USA \\
    $^{2}$Department of Liberal Arts and Sciences, \\
    Virginia Commonwealth University School of the Arts in Qatar, Doha, Qatar
  }
  \maketitle
} \fi

\if1\blind
{
  \bigskip
  \bigskip
  \bigskip
  \begin{center}
    {\LARGE\bf Title}
\end{center}
  \medskip
} \fi

\bigskip

\begin{abstract}

Fractional Brownian motion has been widely used in financial modeling to
capture long-range dependence and persistent behavior observed in asset
dynamics. In the fractional Black--Scholes framework, accurate estimation
of the Hurst parameter is essential, since estimation uncertainty can
directly affect option pricing results. In this paper, we propose a
Bayesian framework for joint inference on the Hurst parameter and
volatility in fractional stochastic differential equation models.
In contrast to approaches based solely on point estimation, the proposed method propagates posterior uncertainty directly into option pricing
distributions under the fractional Black--Scholes model.

Simulation studies are conducted across multiple values of the Hurst
parameter and sample sizes to evaluate estimation accuracy, posterior
coverage, and pricing uncertainty. The results demonstrate stable
posterior inference and coherent uncertainty quantification for both
model parameters and option prices. The methodology is further
illustrated using WTI crude oil and natural gas data under different
market regimes. The empirical analysis indicates that differences in
market behavior are driven primarily by changes in volatility rather than
strong long-range dependence, while posterior option price distributions
reflect substantial variation in pricing uncertainty across regimes.
These findings highlight the importance of incorporating joint parameter
uncertainty in fractional financial models and demonstrate the practical
value of Bayesian methods for option pricing applications.

\end{abstract}

\noindent%

\spacingset{1.45}
\section{Introduction}
\label{sec:intro}

The pricing of financial derivatives remains a central topic in mathematical
finance, with the Black--Scholes model representing a fundamental breakthrough
in the field \citep{BlackScholes1973, Merton1973}. By modeling asset prices
through geometric Brownian motion under a no-arbitrage framework, the model
provides a closed-form solution for European option pricing and has become a
cornerstone of modern financial theory. Despite its widespread use, empirical
studies have shown that financial time series often exhibit features that are
inconsistent with the model assumptions, including long-range dependence, volatility clustering, and heavy-tailed return distributions.

Fractional Brownian motion directly addresses these shortcomings by introducing
memory into the stochastic dynamics of asset prices. Fractional Brownian motion
(fBm) was formally introduced by \citet{Mandelbrot1968}, who defined it as a
generalization of standard Brownian motion with stationary, correlated increments.
Its relevance to financial time series---particularly the presence of long-range
dependence---was subsequently argued by \citet{Mandelbrot1971}. Fractional
Brownian motion incorporates a Hurst parameter that characterizes long-range
dependence in stochastic processes. In this framework, the Hurst parameter
$H \in (0,1)$ governs the degree of dependence in the process, while $\sigma > 0$
denotes the volatility. In particular, when the Hurst parameter exceeds one-half,
the process exhibits persistence, reflecting the tendency of financial returns to
display long memory. This has led to the development of fractional Black--Scholes
models, which extend the classical framework by incorporating memory effects and
providing a more flexible representation of asset dynamics \citep{Biagini2008}.
More recent developments in financial modeling have also emphasized rough
volatility and fractional dynamics as realistic representations of market
behavior, particularly at high frequencies \citep{Gatheral2018, Bennedsen2022}.
Several studies have proposed explicit formulations of such models
based on fractional Brownian motion, illustrating how the Hurst
parameter influences option pricing and volatility dynamics
\citep{ComteRenault1998,HuOksendal2003,Njomen2019}.

Early studies also investigated option pricing in markets driven by
fractional Brownian motion and demonstrated the potential impact of
long-range dependence on derivative valuation
\citep{Necula2002}. Notwithstanding these advances, the problem
of jointly estimating the Hurst parameter $H$ and the volatility $\sigma$ within
a unified statistical framework has received comparatively little attention. In
particular, Bayesian approaches that simultaneously quantify uncertainty in both
parameters remain underdeveloped, despite their natural suitability for this
inference problem.

While fractional models offer improved realism, their practical implementation
introduces significant statistical challenges. In particular, the Hurst index
and volatility must be inferred from observed data, and their joint behavior
directly influences pricing outcomes. Many existing approaches rely on point
estimates or marginal intervals, which may fail to adequately capture the
dependence between these quantities. As a result, estimation error can be
misrepresented, potentially leading to inaccurate assessments in financial
applications. From a practical perspective, these challenges are especially
important in decision-making, where even small inaccuracies can lead to
meaningful differences in option values and affect hedging strategies, risk
management, and portfolio allocation.

Accurate estimation of the Hurst parameter has been a central challenge in the application of fractional models to financial data. The statistical theory of long-range dependence has been extensively developed in both Gaussian and non-Gaussian settings, with applications
across telecommunications, hydrology, economics, and finance \citep{Beran1994, EmbrechtsMaejima2002, Doukhan2003, SamorodnitskyTaqqu1994}. These developments have established the Hurst parameter as a key measure of persistence and dependence in stochastic processes.

A wide range of methods has been proposed for estimating long-range dependence, each with different trade-offs between bias, efficiency, and computational complexity. Early work by \citet{Beran1994} provided a systematic treatment of statistical methods for long-memory processes, while subsequent studies highlighted the sensitivity of classical estimators to short-range dependence and finite-sample effects \citep{Taqqu1995}. Semi-parametric approaches, such as log-periodogram regression and local Whittle estimation, have been widely used because of their attractive asymptotic properties, although they often require relatively large sample sizes to achieve reliable performance \citep{Robinson1995, Hurvich1998}. Alternative approaches include wavelet-based estimators, which exploit the scale invariance of fractional processes and have been shown to improve estimation accuracy in certain settings \citep{Abry1998, Bardet2000}. A broader overview of commonly used Hurst exponent estimation procedures and their practical characteristics is provided by \citet{Zhang2024}.

Despite these developments, much of the existing literature focuses primarily on point estimation and does not fully account for uncertainty in the estimated parameters. This limitation is particularly important in financial applications, where small errors in estimating the Hurst parameter and volatility can propagate through pricing formulas and lead to substantial differences in option valuation. These considerations motivate the use of Bayesian approaches, which provide full posterior distributions and allow uncertainty in model parameters to be carried directly into subsequent inference and pricing decisions.

More recently, Bayesian approaches have been proposed as a flexible framework for
estimation in long-memory models, allowing for the incorporation of prior
information and providing full posterior distributions \citep{Weron2002,Makarava2012,Beskos2015,Chen2017,Dlask2017,Tsionas2021}.
These methods offer a natural way to quantify uncertainty through posterior
distributions. In many applications, however, inference is primarily
reported through marginal posterior summaries such as posterior means and
credible intervals \citep{Jacquier1994,Makarava2012,Chopin2020}.

At the same time, maximum likelihood and method of moments estimators are known
to exhibit systematic bias when applied to fractional stochastic differential equation models, particularly for values of the Hurst parameter away from 0.5.
This issue has been documented in both parametric and moment-based settings
\citep{GhanamBooneX2024, Sagor2025}, where it is shown that both maximum likelihood and method of moments estimators require bias correction for reliable practical use.

Recent work has also considered Bayesian modeling of static and dynamic
Hurst parameters under stochastic volatility frameworks, further
highlighting the importance of jointly modeling persistence and volatility
in long-memory processes \citep{Tsionas2021}. Despite these developments, the joint dependence between key parameters, such as the Hurst parameter and volatility, is often not fully captured, even though this dependence has a direct impact on derived quantities such as option prices.
In particular, while bias-corrected estimators improve point estimation, they do
not address the propagation of joint parameter uncertainty into financial quantities.
This limitation is especially important in financial applications, where interactions
between parameters can significantly influence pricing outcomes. These observations
highlight the need for methods that explicitly account for joint parameter uncertainty
in fractional models.

Bayesian inference provides a coherent framework for addressing these challenges
by combining prior information with observed data to produce a full posterior
distribution over model parameters \citep{Murphy2012}. A key advantage of the Bayesian approach is its ability to capture dependence between parameters through
joint inference. In this context, joint highest-density regions (HDRs) offer a
principled method for summarizing uncertainty in a multidimensional setting,
providing a more accurate representation than separate marginal intervals.

Motivated by these considerations, this paper focuses on Bayesian estimation of
the Hurst parameter and volatility within a fractional Gaussian noise framework.
Using a simulation-based approach, we construct empirical joint highest-density
regions for $(H,\sigma)$ and evaluate their repeated-sampling performance. The
results reveal strong posterior dependence between parameters and highlight the
limitations of marginal interval-based inference.

To assess the practical implications of parameter uncertainty, posterior samples
are propagated through the fractional Black--Scholes pricing formula to obtain
distributions of option prices. This approach allows for the construction of
credible intervals that reflect joint parameter uncertainty, providing a more
informative basis for decision-making compared to traditional point estimates.

In this paper, we address this gap by developing a Bayesian framework for the
joint estimation of $H$ and $\sigma$ from discretely observed asset price data
following a fractional Black--Scholes model. We construct prior distributions
informed by the theoretical constraints on each parameter, derive the
corresponding posterior, and employ Markov chain Monte Carlo (MCMC) methods for
inference. The resulting estimates are then applied to fractional option pricing,
where we demonstrate that incorporating posterior uncertainty in $H$ and $\sigma$
leads to materially different option prices compared to approaches that treat
these parameters as fixed or known. The remainder of the paper is organized as
follows. Section~\ref{sec:simulation_bayes_hdr} presents the fractional Black--Scholes
framework and the Bayesian methodology used to jointly estimate $H$ and $\sigma$,
including prior specification, posterior computation, and the construction of
joint highest-density regions. Section~\ref{sec:results} reports a simulation
study evaluating the performance of the proposed estimators, with emphasis on
coverage, posterior dependence, and estimation accuracy. Section~\ref{sec:application}
provides an empirical application in which posterior parameter draws are
propagated through the fractional Black--Scholes pricing formula to quantify
option pricing uncertainty. Section~\ref{sec:conclusion} concludes the paper
and discusses limitations and directions for future research.

\section{Methodology}
\label{sec:simulation_bayes_hdr}

This section describes the simulation framework used to study Bayesian estimation of the Hurst parameter $H$ and the scale parameter $\sigma$ under fractional Gaussian noise (fGn). We construct joint highest-density regions (HDRs) for $(H,\sigma)$ and assess repeated-sampling coverage with respect to the true pair $(H_c,\sigma_0)$.

\subsection{Data-generating mechanism}

Let $d=(d_1,\dots,d_n)^\top$ denote an observed fGn sample of length $n$. For a fixed Hurst parameter $H\in(1/2,1)$, the lag-$k$ autocovariance of unit-variance fGn is
\begin{equation}
\rho_H(k)
=
\frac{1}{2}\left(|k+1|^{2H}-2|k|^{2H}+|k-1|^{2H}\right),
\qquad k=0,1,2,\dots.
\label{eq:fgn_acf}
\end{equation}
Since $\rho_H(0)=1$, $\rho_H(k)$ also equals the lag-$k$ autocorrelation, as derived from fractional Brownian motion \citep{Mandelbrot1968, Beran1994}.

Using \eqref{eq:fgn_acf}, the $n\times n$ Toeplitz correlation matrix $R(H)$ is defined by

\begin{equation}
R(H)
=
\begin{pmatrix}
\rho_H(0) & \rho_H(1) & \cdots & \rho_H(n-1)\\
\rho_H(1) & \rho_H(0) & \cdots & \rho_H(n-2)\\
\vdots & \vdots & \ddots & \vdots\\
\rho_H(n-1) & \rho_H(n-2) & \cdots & \rho_H(0)
\end{pmatrix}.
\label{eq:toeplitz_corr}
\end{equation}

This Toeplitz structure is standard in long-memory Gaussian processes
\citep{Beran1994} and provides the basis for data generation in our
simulation study. For fixed true values $(\mu_0,\sigma_0, H_c)$, the simulated sample is generated from the corresponding Gaussian model
using this covariance structure.

\begin{equation}
d \sim N\!\left(\mu_0 \mathbf{1}_n,\; \sigma_0^2 R(H_c)\right),
\label{eq:true_model}
\end{equation}
where $\mathbf{1}_n$ denotes the $n$-vector of ones.

To simulate exactly from \eqref{eq:true_model}, we compute the Cholesky factorization
\begin{equation}
R(H_c)=U^\top U,
\label{eq:chol_true}
\end{equation}
generate $z\sim N(0,I_n)$, and set
\begin{equation}
d = \mu_0 \mathbf{1}_n + \sigma_0 U^\top z.
\label{eq:sim_sample}
\end{equation}

This yields an exact Gaussian sample with mean $\mu_0\mathbf{1}_n$
and covariance $\sigma_0^2R(H_c)$. Cholesky factorization provides a numerically stable and computationally efficient approach for simulating Gaussian vectors with positive-definite covariance matrices \citep{Golub2013}.

\subsection{Bayesian Model Specification}

Conditional on $(\mu,\sigma,H)$, the data are modeled as
\begin{equation}
d \mid \mu,\sigma,H \sim N\!\left(\mu \mathbf{1}_n,\; \sigma^2 R(H)\right),
\label{eq:model_likelihood}
\end{equation}
with $H\in(H_L,H_U)$, where in our implementation $H_L=0.5$ and $H_U=1$.

The prior distributions are chosen as
\begin{align}
\mu &\sim N(\mu_0^\ast,\tau_\mu^2), \label{eq:prior_mu}\\
\sigma &\sim \mathrm{Lognormal}(m_\sigma,s_\sigma^2), \label{eq:prior_sigma}\\
H &\sim \mathrm{Uniform}(H_L,H_U), \label{eq:prior_H}
\end{align}

where $\mu_0^\ast$, $\tau_\mu^2$, $m_\sigma$, and $s_\sigma^2$ are fixed hyperparameters, following standard Bayesian modeling practice \citep{Murphy2012}.

Hence, the posterior density is
\begin{equation}
\pi(\mu,\sigma,H\mid d)
\propto
L(d\mid \mu,\sigma,H)\,
\pi(\mu)\pi(\sigma)\pi(H),
\label{eq:posterior_basic}
\end{equation}
where $L(d\mid \mu,\sigma,H)$ is the multivariate normal likelihood induced by \eqref{eq:model_likelihood}.

\subsection{Posterior Computation}

Under \eqref{eq:model_likelihood}, the log-likelihood is
\begin{align}
\log L(d\mid \mu,\sigma,H)
=
-\frac{1}{2}\Big[
&
n\log(2\pi)
+n\log(\sigma^2)
+\log |R(H)|
\nonumber\\
&\quad
+\frac{1}{\sigma^2}(d-\mu\mathbf{1}_n)^\top R(H)^{-1}(d-\mu\mathbf{1}_n)
\Big].
\label{eq:loglik}
\end{align}

which follows from the multivariate Gaussian likelihood \citep{Beran1994}.
To compute this efficiently, we use the Cholesky factorization
\begin{equation}
R(H)=C(H)^\top C(H),
\label{eq:chol_H}
\end{equation}
so that
\begin{equation}
\log |R(H)| = 2\sum_{i=1}^n \log C_{ii}(H),
\label{eq:logdet}
\end{equation}
and the quadratic form is evaluated through triangular solves rather than direct matrix inversion, which improves numerical stability and computational efficiency.

Combining \eqref{eq:loglik} with the priors \eqref{eq:prior_mu}--\eqref{eq:prior_H}, the log-posterior is
\begin{equation}
\log \pi(\mu,\sigma,H\mid d)
=
\log L(d\mid \mu,\sigma,H)
+\log \pi(\mu)
+\log \pi(\sigma)
+\log \pi(H)
+ \text{constant}.
\label{eq:logposterior}
\end{equation}

\subsection{Bayesian Implementation}

To facilitate efficient MCMC sampling and avoid boundary issues, we work with
transformed parameters \citep{Chopin2020}. For the volatility parameter
$\sigma>0$, we define $\eta=\log\sigma$. For the Hurst parameter
$H\in(H_L,H_U)$, we apply a logit transformation
\begin{equation}
z_H = \log\left(\frac{H-H_L}{H_U-H_L - (H-H_L)}\right),
\end{equation}
which maps $H$ to the real line. The corresponding Jacobian adjustments are
included in the posterior target density. 

Posterior inference is carried out using a Metropolis-within-Gibbs scheme with random-walk proposals \citep{Chopin2020}. To improve sampling efficiency, we incorporate a joint block update for $(H,\sigma)$, motivated by their strong posterior dependence. The proposal covariance is estimated from a preliminary pilot run and used to construct a bivariate Gaussian random-walk proposal aligned with the posterior geometry. This substantially improves mixing in regions of high posterior curvature.

\subsection{Joint Highest-Density Region Coverage}

Following the posterior sampling step, we evaluate the transformed posterior
log-density at each draw as
\begin{equation}
h^{(m)} = \log \widetilde{\pi}(\mu^{(m)},\eta^{(m)},z_H^{(m)}\mid d),
\label{eq:height_each_draw}
\end{equation}
which is used to construct the joint highest-density regions.

The transformed posterior log-density is evaluated at each retained
posterior draw $m=1,\ldots,M$, producing a set of posterior heights
$\{h^{(m)}\}_{m=1}^M$. These values are ranked from largest to smallest.
For a target probability level $\alpha \in \{0.90,0.95\}$, we retain the
top $\alpha M$ posterior draws following the highest-density region
(HDR) construction of \citep{Hyndman1996}. The retained subset defines
an empirical joint HDR for $(H,\sigma)$:
\begin{equation}
\mathcal{R}_\alpha
=
\left\{
(H^{(m)},\sigma^{(m)}): 
\log \widetilde{\pi}(\mu^{(m)},\eta^{(m)},z_H^{(m)}\mid d)
\ge c_\alpha
\right\},
\label{eq:hdr_region}
\end{equation}
where $c_\alpha$ is the empirical cutoff corresponding to the top $\alpha$ fraction of posterior heights. This construction is preferable to separate marginal intervals for $H$ and $\sigma$, since the posterior draws exhibit substantial dependence and lie on a curved ridge rather than in an approximately rectangular region.


To evaluate repeated-sampling performance, we repeat the above procedure over $B$ simulated datasets. For each replication, we check whether the true parameter pair $(H_c,\sigma_0)$ lies inside the estimated joint HDR:
\begin{equation}
I_b
=
\mathbf{1}\left\{
(H_c,\sigma_0)\in \mathcal{R}_\alpha^{(b)}
\right\},
\qquad b=1,\dots,B.
\label{eq:cover_indicator}
\end{equation}

The empirical coverage is then
\begin{equation}
\widehat{\mathrm{Cov}}_\alpha
=
\frac{1}{B}\sum_{b=1}^B I_b.
\label{eq:emp_coverage}
\end{equation}

In addition, we summarize posterior medians, biases and posterior correlations between $H$ and $\sigma$.

\subsection{Posterior Option Pricing}

To connect parameter uncertainty with option pricing, each retained
posterior draw $(H^{(m)},\sigma^{(m)})$ is mapped through the
fractional Black--Scholes pricing framework
\citep{HuOksendal2003,Biagini2008,Njomen2019}. The pricing equation is used here as a
sensitivity-based mechanism for propagating posterior uncertainty in
$(H,\sigma)$ into option values. Let
\begin{equation}
\lambda_H = 2H T^{2H-1}.
\label{eq:lambda_H}
\end{equation}
Then
\begin{align}
d_1 &=
\frac{\log(S_0/K)+\left(r+\frac{\lambda_H}{2}\sigma^2\right)T}
{\sigma\sqrt{\lambda_H T}},
\label{eq:d1_fbsm}
\\
d_2 &=
\frac{\log(S_0/K)+\left(r-\frac{\lambda_H}{2}\sigma^2\right)T}
{\sigma\sqrt{\lambda_H T}},
\label{eq:d2_fbsm}
\end{align}
and the call price is
\begin{equation}
C(H,\sigma)
=
S_0 \Phi(d_1) - K e^{-rT}\Phi(d_2),
\label{eq:fbsm_call}
\end{equation}
where $\Phi(\cdot)$ denotes the standard normal distribution function.

Applying \eqref{eq:fbsm_call} to every posterior draw produces a posterior sample of option prices,
\begin{equation}
C^{(m)} = C(H^{(m)},\sigma^{(m)}), \qquad m=1,\dots,M,
\label{eq:post_call_draws}
\end{equation}
from which posterior medians and credible intervals for the option price are obtained.

\section{Simulation Study}
\label{sec:results}

This section evaluates the repeated-sampling performance of the proposed
Bayesian procedure under fractional Gaussian noise. For each replication, we (i) generate an exact Gaussian fGn sample from the true covariance structure, (ii) estimate $(\mu,\sigma,H)$ via Bayesian MCMC under the multivariate normal model, (iii) construct an empirical joint HDR by ranking posterior draws according to posterior height, (iv) record whether the true pair $(H_c,\sigma_0)$ falls inside the estimated HDR, and (v) propagate posterior draws through the fractional Black--Scholes formula to obtain a posterior distribution for the call price.

Posterior inference was based on 3,000 MCMC iterations following an
initial pilot run of 1,000 iterations. The first 500 iterations of the
main chain were discarded as burn-in, leaving 2,500 posterior draws for
inference. For each design point, 100 simulated datasets were generated.
Computation times are based on a MacBook Air with an Apple M4 processor
and 24GB of RAM. Approximate runtimes for the full simulation grid under
the sequential implementation were 6.6 minutes for $n=50$, 16.9 minutes
for $n=100$, and 4.5 hours for $n=300$, demonstrating the substantial
increase in computational cost as sample size grows. The largest setting,
$n=500$, required approximately three days under the sequential
implementation. To improve computational efficiency, a parallelized version
of the simulation procedure was also implemented, reducing the runtime for
$n=500$ to approximately four hours while producing qualitatively
consistent results. 

Across all simulation settings, estimation accuracy improved as the
sample size increased, with narrower posterior regions and reduced bias
observed for larger values of $n$. However, the overall pricing behavior
and uncertainty structure remained qualitatively similar for $n \geq 300$.
This suggests that moderate sample sizes already provide stable inference
for the proposed Bayesian framework while avoiding the substantially
higher computational cost associated with larger datasets. Similar
computational trade-offs have been noted in recent studies of Bayesian
Hurst exponent estimation for long-memory models
\citep{Mangalam2025}.

\begin{figure}[htbp]
\centering
\includegraphics[width=\textwidth]{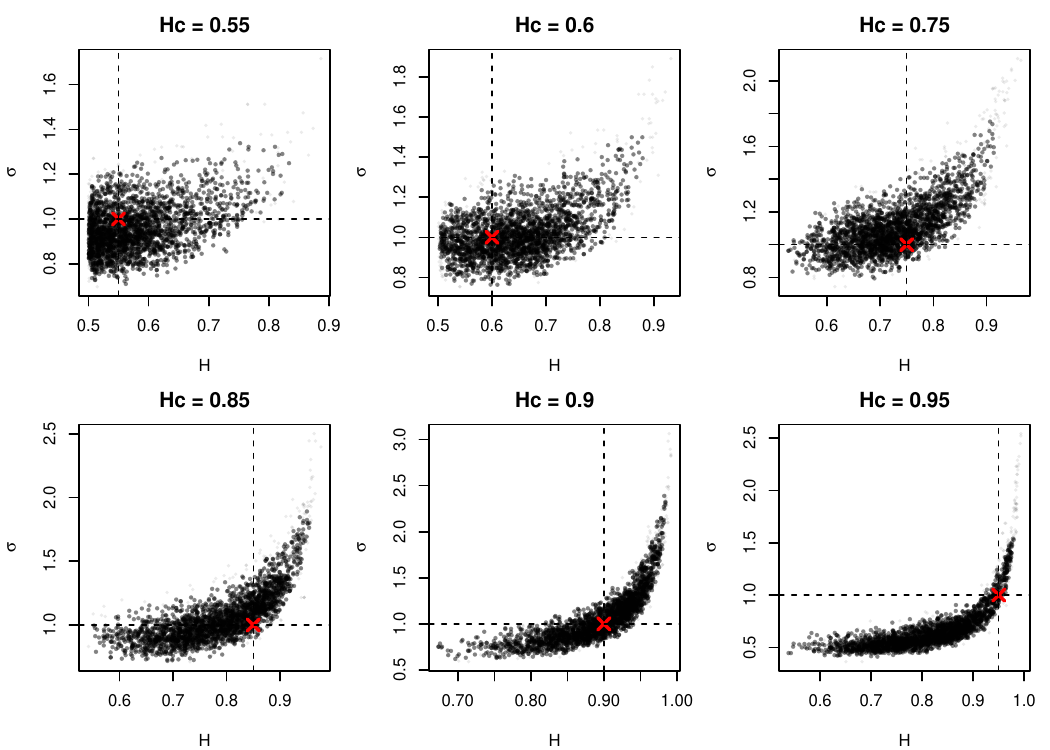}
\caption{Joint posterior draws and 95\% HDR cloud for $n=50$ (gray: posterior draws; darker: top 95\% HDR; red marker: true $(H_c,\sigma_0)$).}
\label{fig:hdr_n50}
\end{figure}

\begin{figure}[htbp]
\centering
\includegraphics[width=\textwidth]{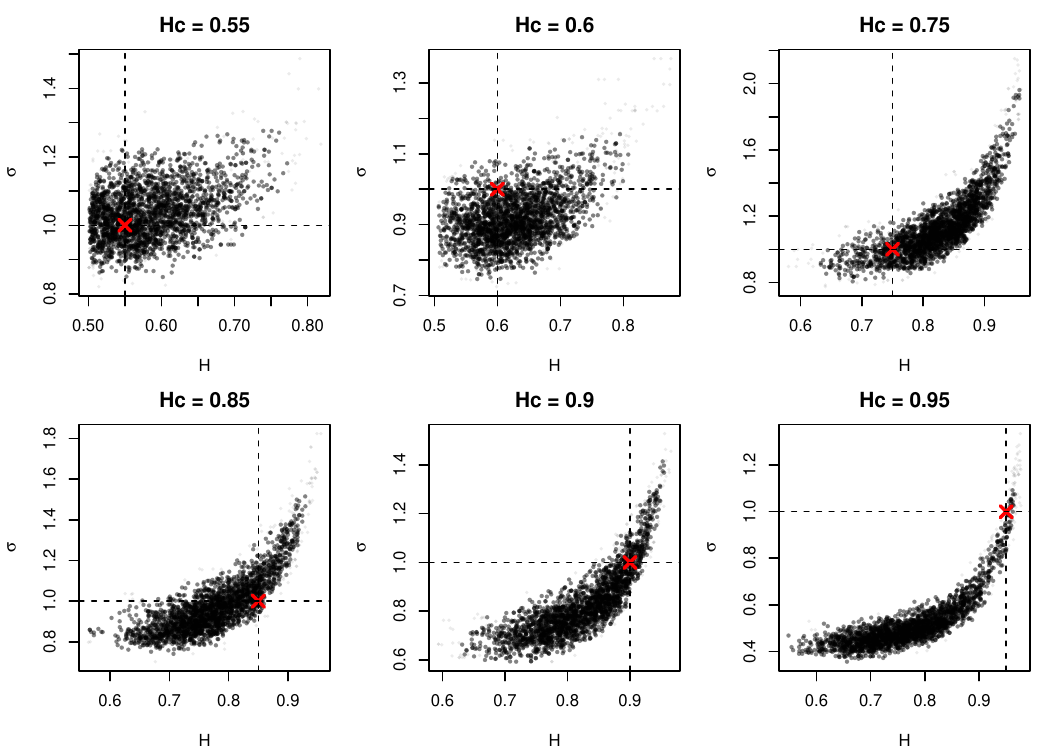}
\caption{Joint posterior draws and 95\% HDR cloud for $n=100$ (gray: posterior draws; darker: top 95\% HDR; red marker: true $(H_c,\sigma_0)$).}
\label{fig:hdr_n100}
\end{figure}

\begin{figure}[htbp]
\centering
\includegraphics[width=\textwidth]{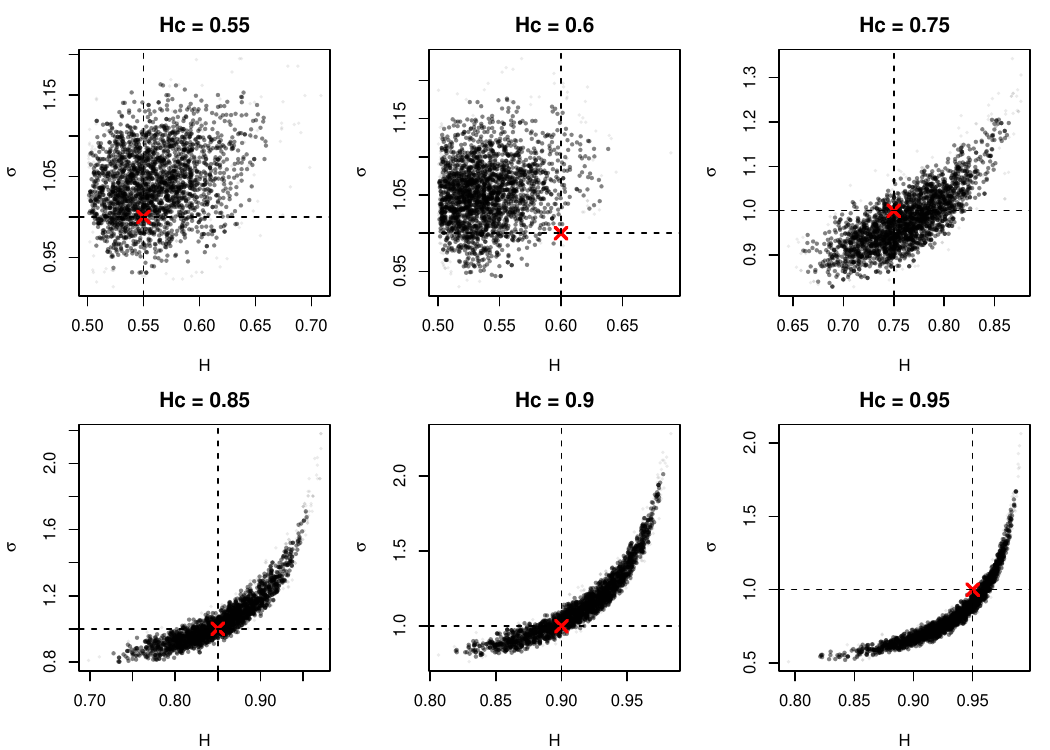}
\caption{Joint posterior draws and 95\% HDR cloud for $n=300$ (gray: posterior draws; darker: top 95\% HDR; red marker: true $(H_c,\sigma_0)$).}
\label{fig:hdr_n300}
\end{figure}

\begin{figure}[htbp]
\centering
\includegraphics[width=\textwidth]{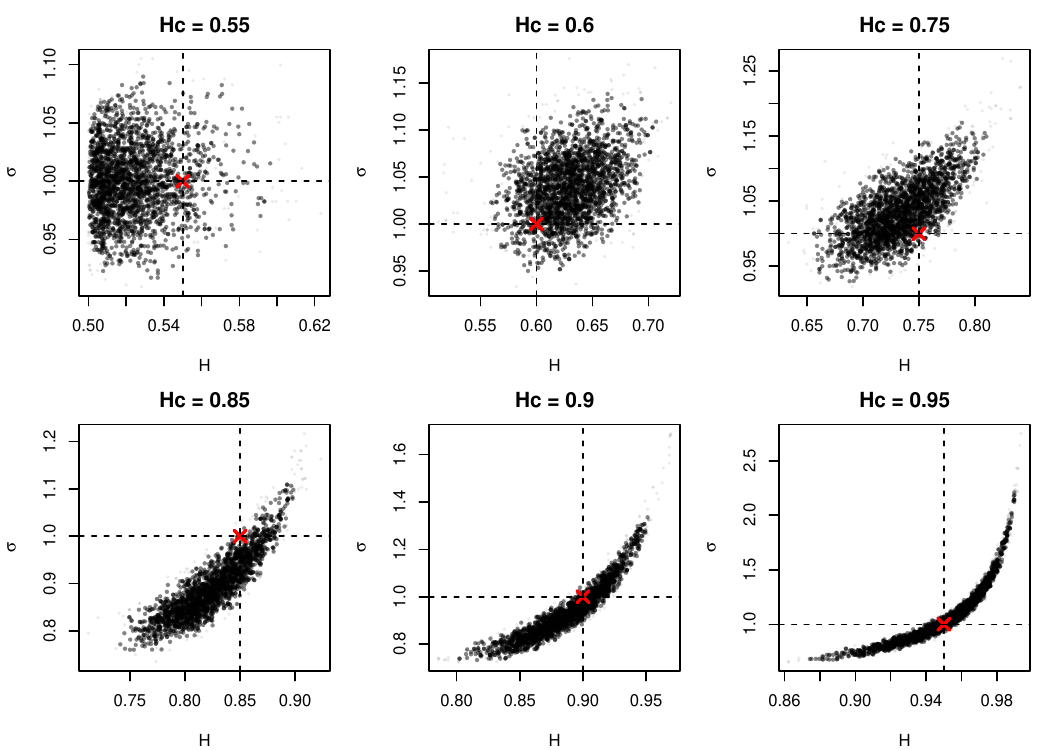}
\caption{Joint posterior draws and 95\% HDR cloud for $n=500$ (gray: posterior draws; darker: top 95\% HDR; red marker: true $(H_c,\sigma_0)$).}
\label{fig:hdr_n500}
\end{figure}

\begin{figure}[htbp]
\centering
\begin{subfigure}{0.7\textwidth}
  \centering
  \includegraphics[width=\textwidth]{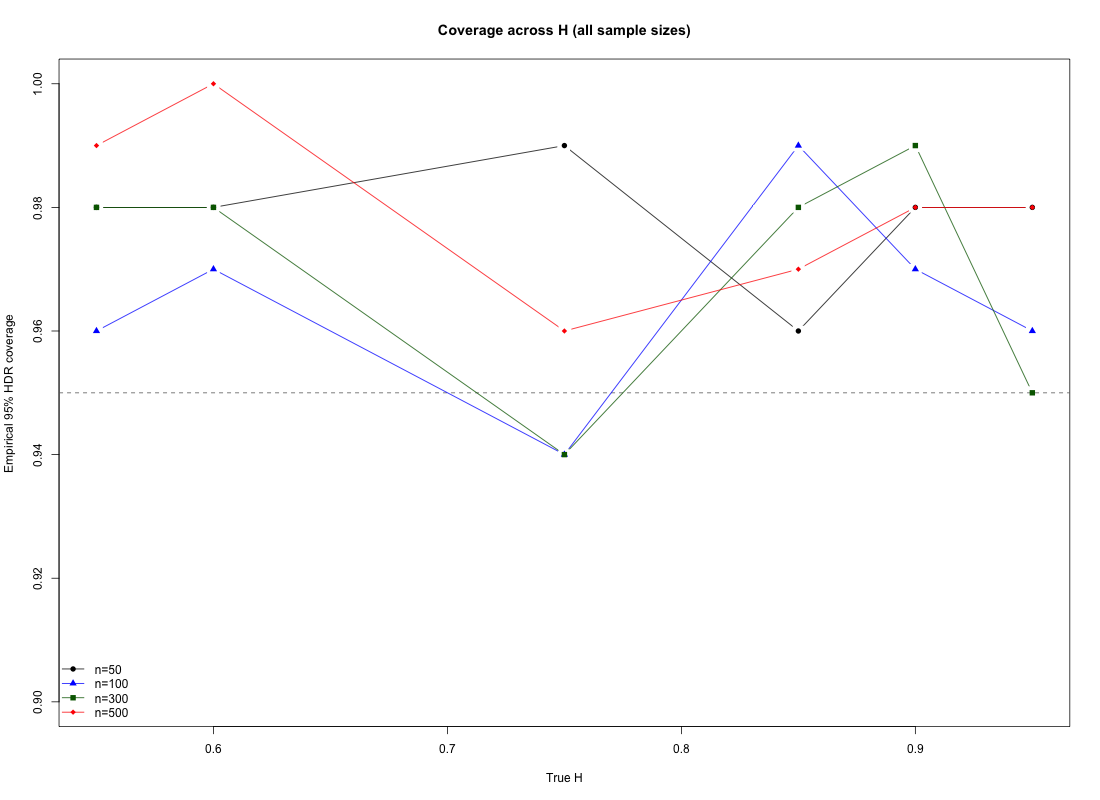}
  \caption{Empirical 95\% HDR coverage across $H$ (all sample sizes). The dashed line indicates the nominal 0.95 level.}
  \label{fig:alln_cov}
\end{subfigure}
\hfill

\begin{subfigure}{0.7\textwidth}
  \centering
  \includegraphics[width=\textwidth]{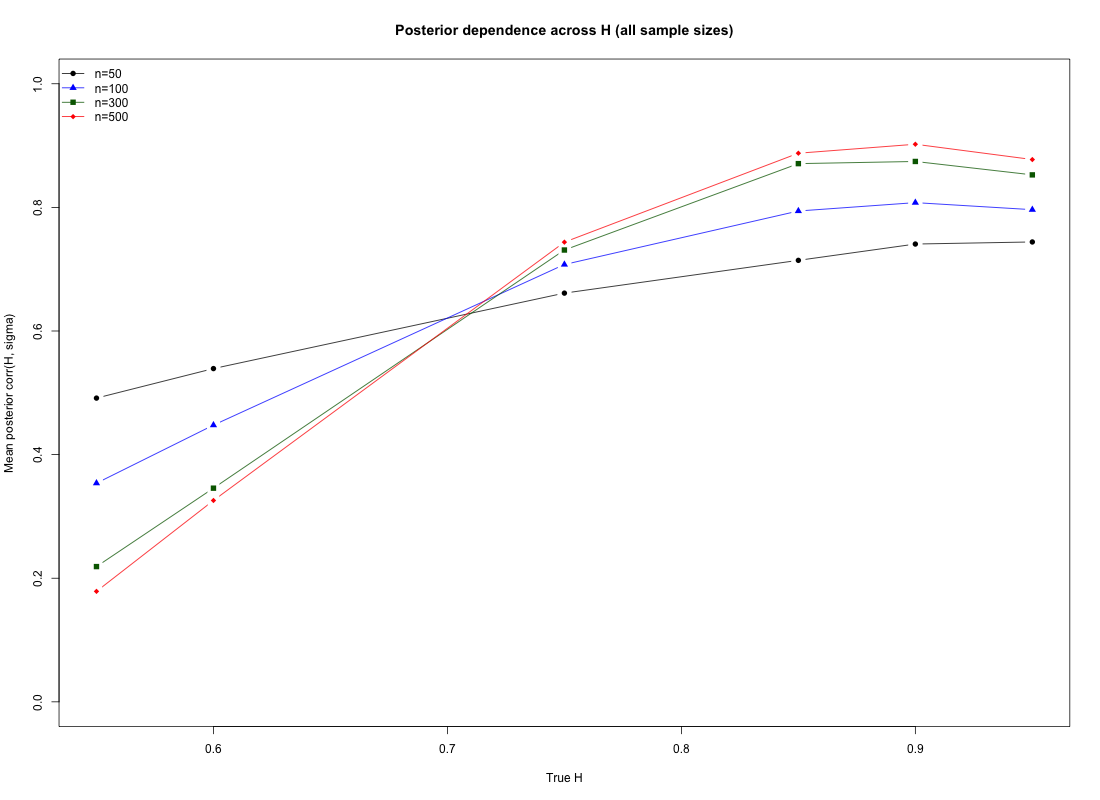}
  \caption{Mean posterior dependence $\mathrm{corr}(H,\sigma)$ across $H$ (all sample sizes).}
  \label{fig:alln_corr}
\end{subfigure}
\caption{Summary panels for the simulation study. Each line corresponds to a different sample size $n \in \{50,100,300,500\}$.}
\label{fig:alln_bc}
\end{figure}

Figures~\ref{fig:hdr_n50}--\ref{fig:hdr_n500} visualize the joint posterior
draws and corresponding 95\% HDR clouds for each sample size. The clouds
highlight pronounced posterior dependence between $H$ and $\sigma$ and
illustrate the resulting non-rectangular uncertainty regions.
Figures~\ref{fig:alln_cov}--\ref{fig:alln_corr} summarize empirical
coverage and posterior dependence across sample sizes on a common scale.
Coverage remains close to the nominal 0.95 level across most settings,
while $\mathrm{corr}(H,\sigma)$ generally increases with $H_c$, indicating
that separate marginal intervals may fail to capture important joint
posterior structure.

It is well known that fractional Brownian motion models may violate the
semimartingale structure underlying classical arbitrage-free pricing
theory \citep{Rogers1997,Biagini2008}. Consequently, the fractional
Black--Scholes framework is used here primarily as a sensitivity-based
tool for studying how posterior uncertainty in $(H,\sigma)$ propagates
into option values, rather than as a complete market equilibrium model
under all fractional specifications. Nevertheless, recent developments in
rough volatility modeling further support the usefulness of fractional
dynamics as flexible representations of financial dependence structures
\citep{Gatheral2018,Bennedsen2022}.

\begin{figure}[htbp]
\centering
\includegraphics[width=\textwidth]{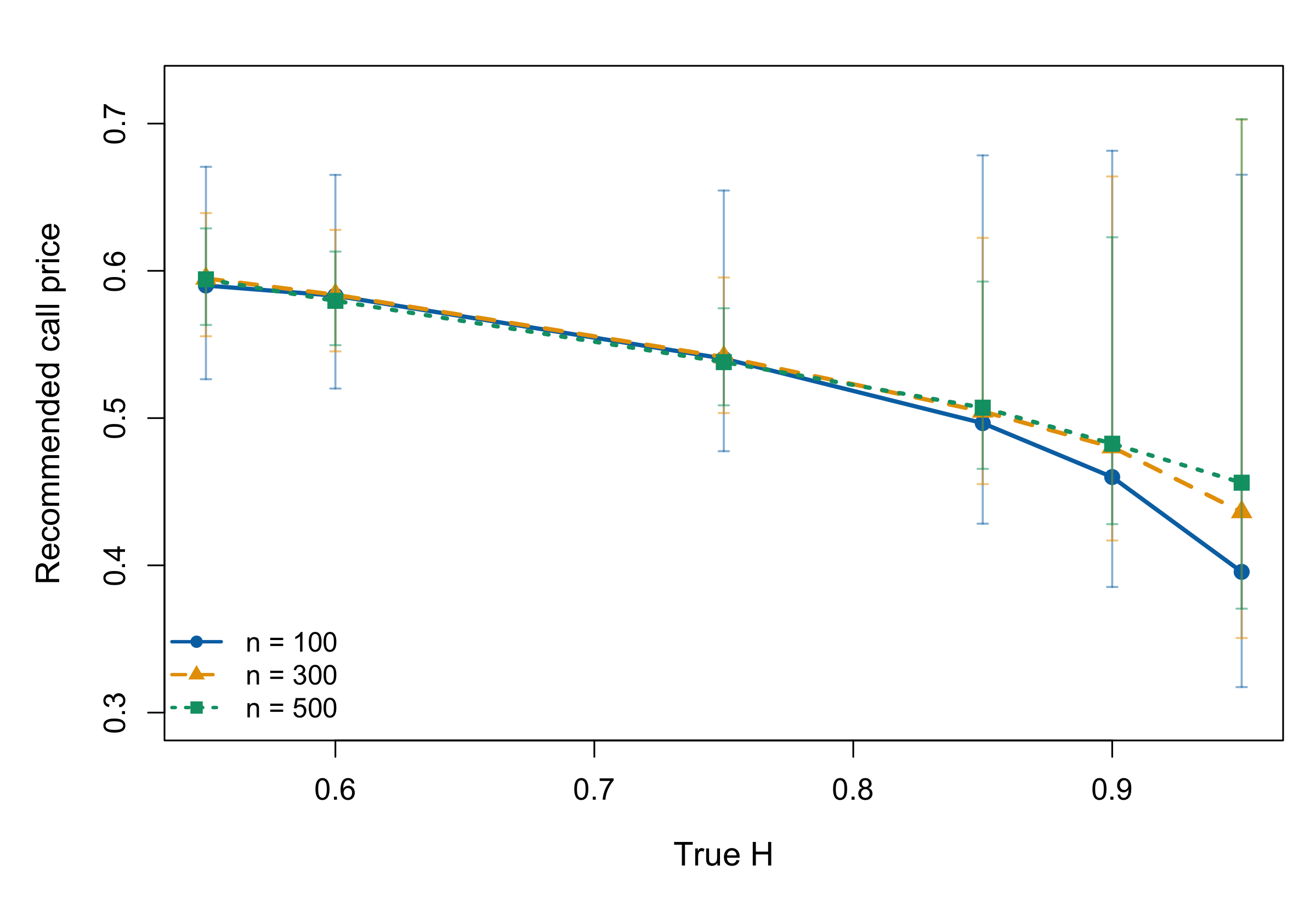}
\caption{Recommended call price (posterior median) versus the true Hurst parameter $H_c$ with 95\% posterior intervals. Curves correspond to sample sizes $n\in\{100,300,500\}$.}
\label{fig:call_alln}
\end{figure}

To quantify the implications for option valuation,
Figure~\ref{fig:call_alln} reports posterior medians and 95\% credible
intervals for the call price obtained by mapping posterior draws
$(H,\sigma)$ through the fractional Black--Scholes pricing equation. Pricing uncertainty decreases
as $n$ increases, consistent with posterior concentration in $(H,\sigma)$,
while the posterior median call price decreases with $H_c$ under the
chosen option parameters. Table~\ref{tab:sim_call_bands} provides the
corresponding numerical summaries, including the 95\% interval
$[q_{0.025},q_{0.975}]$ and the 90\% band
$[q_{0.05},q_{0.95}]$, which may be interpreted as a conservative
bid--ask range.

\begin{table}[htbp]
\centering
\small
\setlength{\tabcolsep}{3pt}
\renewcommand{\arraystretch}{1.1}
\caption{Simulation summary ($B=100$). Coverage is the empirical proportion of replications where the true pair $(H_c,\sigma_0)$ lies inside the estimated 95\% joint HDR. Posterior call-price quantiles are reported in increasing order.}
\label{tab:sim_call_bands}
\begin{adjustbox}{max width=1.15\textwidth,center}
\begin{tabular}{ccccccccccc}
\toprule
$n$ & $H_c$ & Coverage & Mean $H$ Bias & Mean $\sigma$ Bias & Mean Corr$(H,\sigma)$ 
& $C_{0.025}$ & $C_{0.05}$ & $C_{0.5}$ & $C_{0.95}$ & $C_{0.975}$ \\
\midrule
50  & 0.55 & 0.98 &  0.0617 &  0.0213 & 0.4914 & 0.4974 & 0.5097 & 0.5836 & 0.6840 & 0.7085 \\
50  & 0.60 & 0.98 &  0.0379 &  0.0495 & 0.5392 & 0.5033 & 0.5153 & 0.5904 & 0.6947 & 0.7206 \\
50  & 0.75 & 0.99 & -0.0204 &  0.0103 & 0.6613 & 0.4593 & 0.4710 & 0.5447 & 0.6685 & 0.7083 \\
50  & 0.85 & 0.96 & -0.0541 & -0.0550 & 0.7143 & 0.4110 & 0.4215 & 0.4948 & 0.6471 & 0.7001 \\
50  & 0.90 & 0.98 & -0.0542 & -0.0970 & 0.7407 & 0.3755 & 0.3857 & 0.4613 & 0.6427 & 0.7040 \\
50  & 0.95 & 0.98 & -0.0641 & -0.2362 & 0.7440 & 0.3054 & 0.3143 & 0.3901 & 0.6005 & 0.6716 \\
\midrule
100 & 0.55 & 0.94 &  0.0250 &  0.0226 & 0.3538 & 0.5243 & 0.5356 & 0.5955 & 0.6561 & 0.6706 \\
100 & 0.60 & 0.98 &  0.0152 &  0.0265 & 0.4477 & 0.5167 & 0.5293 & 0.5863 & 0.6502 & 0.6652 \\
100 & 0.75 & 0.96 &  0.0007 &  0.0274 & 0.7076 & 0.4413 & 0.4545 & 0.5456 & 0.6590 & 0.6775 \\
100 & 0.85 & 0.98 & -0.0182 & -0.0043 & 0.7941 & 0.4092 & 0.4211 & 0.5044 & 0.6503 & 0.6696 \\
100 & 0.90 & 0.98 & -0.0199 & -0.0314 & 0.8077 & 0.3838 & 0.3956 & 0.4762 & 0.6419 & 0.6664 \\
100 & 0.95 & 0.99 & -0.0341 & -0.1552 & 0.7962 & 0.3056 & 0.3167 & 0.4141 & 0.5901 & 0.6168 \\
\midrule
300 & 0.55 & 0.98 &  0.0081 & -0.0004 & 0.2188 & 0.5534 & 0.5616 & 0.5898 & 0.6318 & 0.6392 \\
300 & 0.60 & 0.98 &  0.0057 &  0.0113 & 0.3457 & 0.5456 & 0.5511 & 0.5832 & 0.6204 & 0.6278 \\
300 & 0.75 & 0.94 & -0.0040 &  0.0072 & 0.7310 & 0.4921 & 0.5007 & 0.5380 & 0.5929 & 0.6012 \\
300 & 0.85 & 0.98 &  0.0006 &  0.0148 & 0.8708 & 0.4496 & 0.4566 & 0.5053 & 0.5814 & 0.5908 \\
300 & 0.90 & 0.99 & -0.0106 & -0.0102 & 0.8743 & 0.4187 & 0.4259 & 0.4810 & 0.5802 & 0.5904 \\
300 & 0.95 & 0.95 & -0.0215 & -0.1101 & 0.8526 & 0.3382 & 0.3461 & 0.4283 & 0.5654 & 0.5866 \\
\midrule
500 & 0.55 & 0.97 &  0.0010 &  0.0036 & 0.1787 & 0.5644 & 0.5680 & 0.5941 & 0.6229 & 0.6288 \\
500 & 0.60 & 0.93 &  0.0050 &  0.0065 & 0.3258 & 0.5510 & 0.5541 & 0.5812 & 0.6074 & 0.6131 \\
500 & 0.75 & 0.96 &  0.0017 &  0.0092 & 0.7437 & 0.4956 & 0.5024 & 0.5371 & 0.5886 & 0.5948 \\
500 & 0.85 & 0.97 & -0.0054 & -0.0052 & 0.8876 & 0.4427 & 0.4487 & 0.4989 & 0.5719 & 0.5800 \\
500 & 0.90 & 0.98 & -0.0065 & -0.0024 & 0.9022 & 0.4256 & 0.4302 & 0.4829 & 0.5693 & 0.5773 \\
500 & 0.95 & 0.98 & -0.0148 & -0.0809 & 0.8774 & 0.3526 & 0.3602 & 0.4379 & 0.5863 & 0.6064 \\
\bottomrule
\end{tabular}
\end{adjustbox}
\end{table}

\begin{figure}[htbp]
\centering
\includegraphics[width=0.9\textwidth]{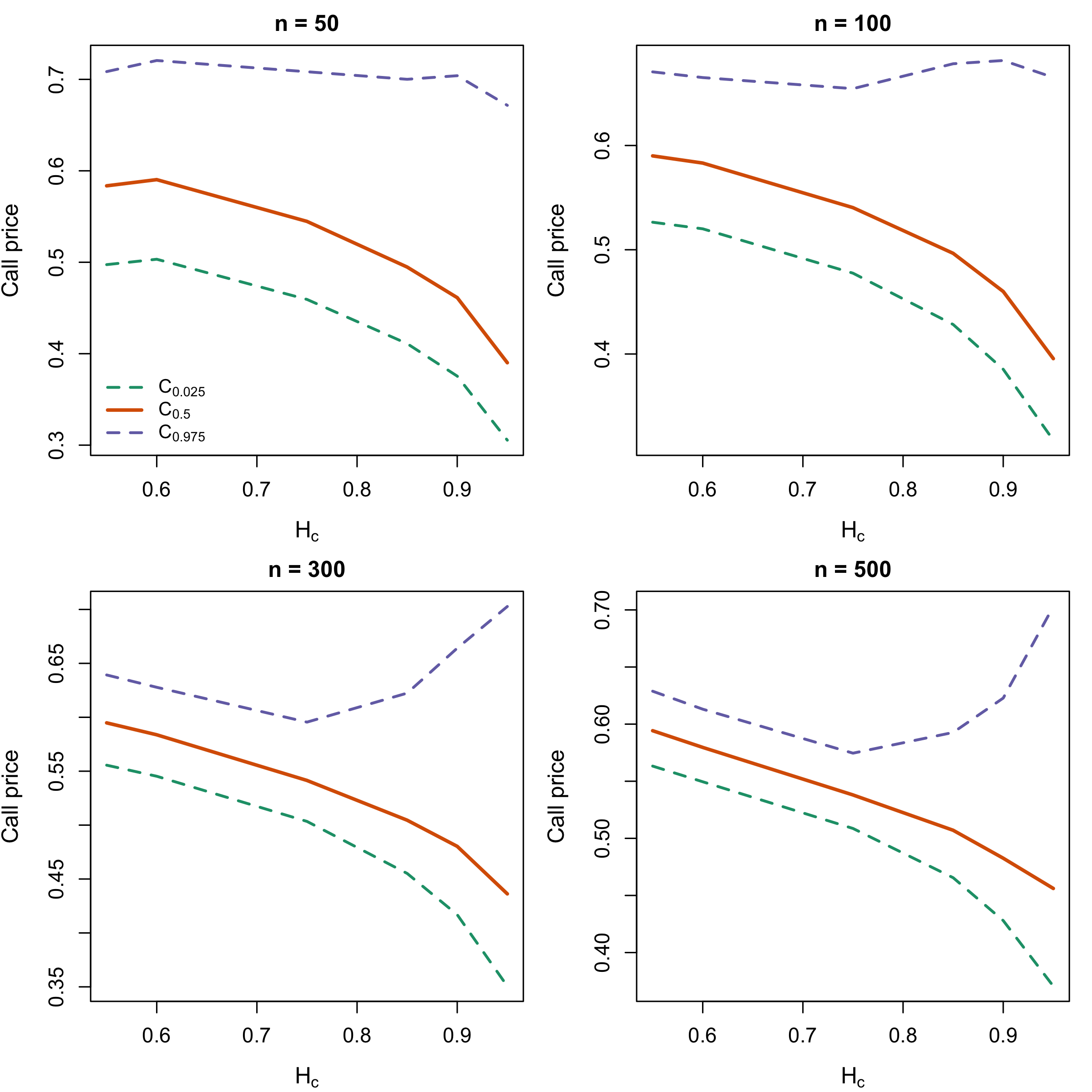}
\caption{Posterior call-price quantiles as functions of the true Hurst parameter $H_c$. 
Each panel corresponds to a different sample size $n \in \{50,100,300,500\}$. 
Solid lines represent the posterior median $C_{0.5}$, while dashed lines represent the lower and upper 95\% credible bounds $C_{0.025}$ and $C_{0.975}$. 
Interval width decreases as $n$ increases, reflecting posterior concentration.}
\label{fig:call_quantiles}
\end{figure}

Figure~\ref{fig:call_quantiles} displays the posterior lower, median,
and upper quantile curves of the option price as functions of the true
Hurst parameter $H_c$ for each sample size. The lower and upper curves form a 95\% credible interval. 
The shrinkage of the gap between $C_{0.025}$ and $C_{0.975}$ as $n$  increases illustrates posterior concentration in the joint parameter 
space. The median call price decreases with $H_c$ under the chosen 
option parameters, while uncertainty is substantially larger for small 
samples and tightens as $n$ increases. These quantile curves also facilitate decision-oriented summaries, 
such as one-sided credible bounds, which may be preferable when 
overpricing and underpricing have asymmetric consequences.

\section{Empirical Application}
\label{sec:application}

This section illustrates the practical performance of the proposed
Bayesian framework using real financial data from energy markets.
We consider daily log returns of WTI crude oil and natural gas futures,
which are known for their pronounced volatility and relevance in
derivative pricing applications.

To investigate the effect of market conditions, the oil data are divided
into two distinct periods representing different volatility regimes.
The first period corresponds to a high-volatility environment, while the
second reflects relatively stable market behavior. Natural gas is included
as an additional example of a persistently volatile asset, allowing us to
examine the robustness of the proposed approach across different settings.

For each dataset, we estimate the Hurst parameter and volatility using
the Bayesian methodology described in the previous section. These estimates
are then incorporated into the fractional Black--Scholes framework to
obtain posterior distributions of option prices. This approach allows us
to assess not only point estimates but also the uncertainty associated
with pricing decisions.

The empirical analysis is designed to distinguish between volatility-driven
effects and genuine long-range dependence, and to quantify their impact on
option pricing. This distinction is particularly important, as traditional
methods may confound volatility with persistence, leading to misleading
inference.

\subsection{WTI Crude Oil: Regime Comparison}

We begin with WTI crude oil, which exhibits distinct regimes of market
behavior. The period 2020--2022 is characterized by elevated volatility,
while 2023--2025 reflects comparatively more stable market conditions.
This provides a natural setting for examining how changes in market
volatility influence posterior inference and option pricing uncertainty.

\begin{table}[H]
\centering
\begin{tabular}{lccccc}
\toprule
Asset & Ticker & Period & $n$ & Mean Return & Annualized Volatility \\
\midrule
WTI Crude Oil & CL=F & 2020--2022 & 522 & 0.00185 & 0.696 \\
WTI Crude Oil & CL=F & 2023--2025 & 752 & $-0.00038$ & 0.314 \\
\bottomrule
\end{tabular}
\caption{Summary statistics for the WTI crude oil return series used in the empirical analysis.}
\label{tab:oil_data_summary}
\end{table}

Table~\ref{tab:oil_data_summary} shows that the earlier period exhibits
substantially higher volatility, while the average returns remain close
to zero in both regimes. This suggests that differences between the two
periods are driven primarily by changes in market variability rather than
systematic shifts in mean returns.

For pricing, we consider a European call option with
$S_0=100$, $K=98$, $T=90/365$, and $r=0.05$, holding these contract
inputs fixed when comparing classical Black--Scholes and fractional
Black--Scholes prices.

\begin{table}[H]
\centering
\begin{tabular}{lccccc}
\toprule
Period & $H$ (95\% CI) & $\sigma$ & BSM & fBSM (95\% CI) & Width \\
\midrule
2020--2022
& 0.532 \,(0.502, 0.586)
& 0.698
& 15.15
& 14.99 \,(14.21, 15.84)
& 1.63 \\

2023--2025
& 0.515 \,(0.501, 0.551)
& 0.315
& 7.86
& 7.82 \,(7.53, 8.12)
& 0.59 \\
\bottomrule
\end{tabular}
\caption{Posterior summaries and option pricing results for WTI crude oil across two market periods. The width column represents the length of the 95\% posterior credible interval for the fractional Black--Scholes option price.}
\label{tab:oil_summary}
\end{table}

Table~\ref{tab:oil_summary} reports posterior summaries for the Hurst
parameter and volatility. Across both periods, the estimated Hurst
parameter remains close to 0.5, indicating weak long-range dependence in
returns. In contrast, posterior volatility estimates differ substantially
between periods, with the 2020--2022 regime exhibiting considerably
higher uncertainty. The wider posterior pricing interval during this
period, with a width 1.63 compared to 0.59 in the later regime, indicates
that elevated volatility translates directly into greater uncertainty in
option valuation.

\begin{figure}[H]
\centering
\includegraphics[width=0.9\textwidth]{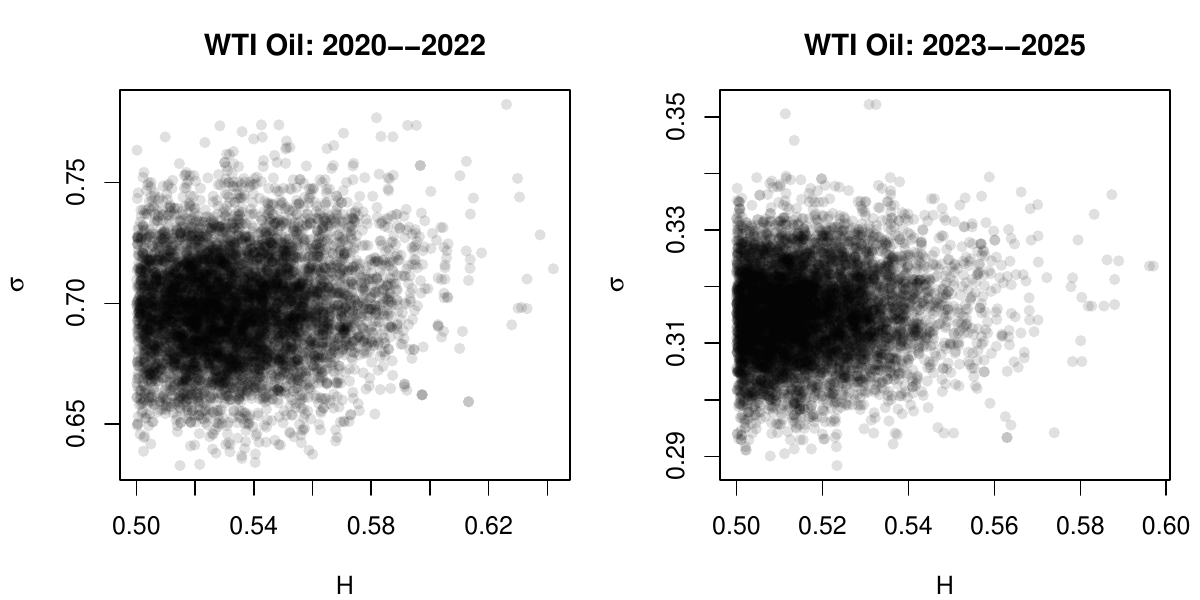}
\caption{Joint posterior distribution of the Hurst parameter $H$ and volatility $\sigma$
for WTI crude oil across two periods. While $H$ remains concentrated around 0.5,
volatility differs substantially between regimes.}
\label{fig:oil_joint}
\end{figure}

Figure~\ref{fig:oil_joint} further illustrates that the posterior
distribution of the Hurst parameter remains relatively stable across
periods, whereas volatility exhibits substantial variation between market
regimes. The posterior clouds also reveal noticeable dependence between
$H$ and $\sigma$, emphasizing the importance of joint uncertainty
quantification in the Bayesian framework. 

The fractional Black--Scholes formula is used here as a sensitivity-based
pricing framework to study how posterior uncertainty in $(H,\sigma)$
propagates into option values, rather than as a claim of complete market
arbitrage-free dynamics under all fractional specifications.

\begin{figure}[H]
\centering
\includegraphics[width=0.9\textwidth]{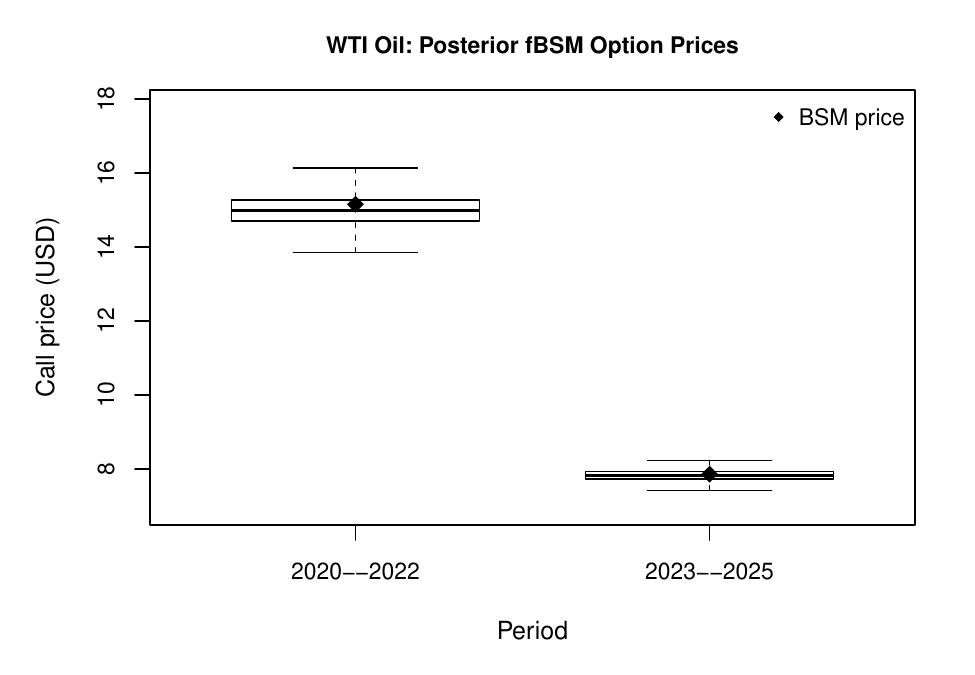}
\caption{Posterior distribution of option prices under the fractional
Black--Scholes model for WTI crude oil across two periods.}
\label{fig:oil_boxplot}
\end{figure}

Figure~\ref{fig:oil_boxplot} shows that the high-volatility period
produces a substantially wider posterior distribution of option prices,
reflecting increased pricing uncertainty. In contrast, the more stable
2023--2025 regime yields a considerably tighter distribution. Although
the posterior median prices remain reasonably close to the corresponding
classical Black--Scholes prices, the range of plausible option values
differs substantially across market regimes. 

Overall, these findings suggest that differences in market regimes affect
option pricing uncertainty primarily through changes in volatility rather
than through strong long-range dependence in returns.

\subsection{Natural Gas: High-Volatility Market}
We next consider natural gas as an example of a persistently high-volatility
commodity market. In contrast to the regime-based behavior observed in
oil, natural gas exhibits consistently high volatility over the entire
sample period. 

\begin{table}[H]
\centering
\begin{tabular}{lccccc}
\toprule
Asset & Ticker & Period & $n$ & Mean Return & Annualized Volatility \\
\midrule
Natural Gas Futures & NG=F & 2022--2024 & 711 & $-0.00041$ & 0.832 \\
\bottomrule
\end{tabular}
\caption{Summary statistics for the natural gas return series used in the empirical analysis.}
\label{tab:ng_data_summary}
\end{table}

For pricing, we consider a European call with $S_0=100$, $K=98$, $T=90/365$, and $r=0.05$, holding these contract inputs fixed when comparing BSM and fBSM prices.

\begin{table}[H]
\centering
\begin{tabular}{lccccc}
\toprule
Period & $H$ (95\% CI) & $\sigma$ & BSM & fBSM (95\% CI) & Width \\
\midrule
2022--2024
& 0.505 \,(0.500, 0.526)
& 0.833
& 17.75
& 17.72 \,(16.94, 18.56)
& 1.62 \\
\bottomrule
\end{tabular}
\caption{Posterior summaries and option pricing results for natural gas. The width column represents the length of the 95\% posterior credible interval for the fractional Black--Scholes option price.}
\label{tab:ng_summary}
\end{table}

Table~\ref{tab:ng_summary} reports posterior summaries of the Hurst
parameter and volatility. The estimated Hurst parameter remains close
to 0.5, indicating weak long-range dependence in returns. In contrast,
the volatility is relatively high, reflecting the pronounced variability
of the natural gas market. The corresponding posterior pricing interval width of 1.62 further indicates substantial uncertainty in option valuation under highly volatile market conditions.

\begin{figure}[H]
\centering
\includegraphics[width=0.75\textwidth]{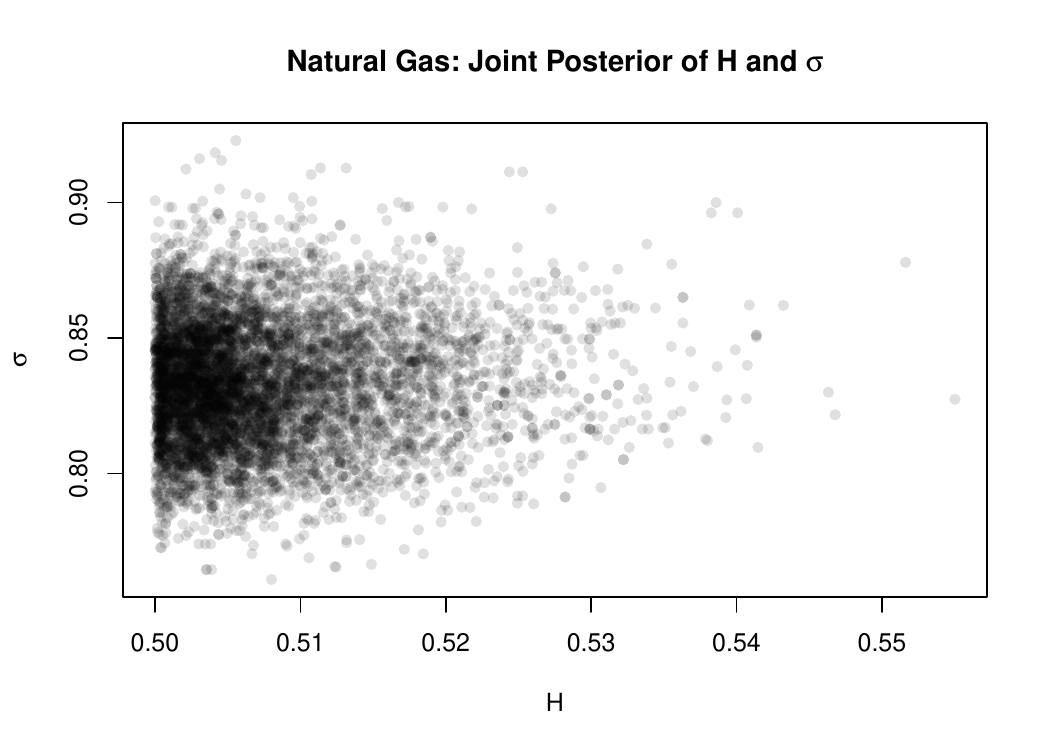}
\caption{Joint posterior distribution of the Hurst parameter $H$ and
volatility $\sigma$ for natural gas. The estimates of $H$ are concentrated
around 0.5, while $\sigma$ exhibits substantial dispersion.}
\label{fig:ng_joint}
\end{figure}

The joint posterior distribution in Figure~\ref{fig:ng_joint} reveals a
concentration of $H$ around 0.5, while the volatility parameter displays
considerable variability. This pattern suggests that the observed dynamics
are primarily driven by volatility rather than persistence. The substantial posterior dispersion in $\sigma$ also contributes directly
to the wider range of plausible option prices observed under the fractional
pricing framework.

\begin{figure}[H]
\centering
\includegraphics[width=0.75\textwidth]{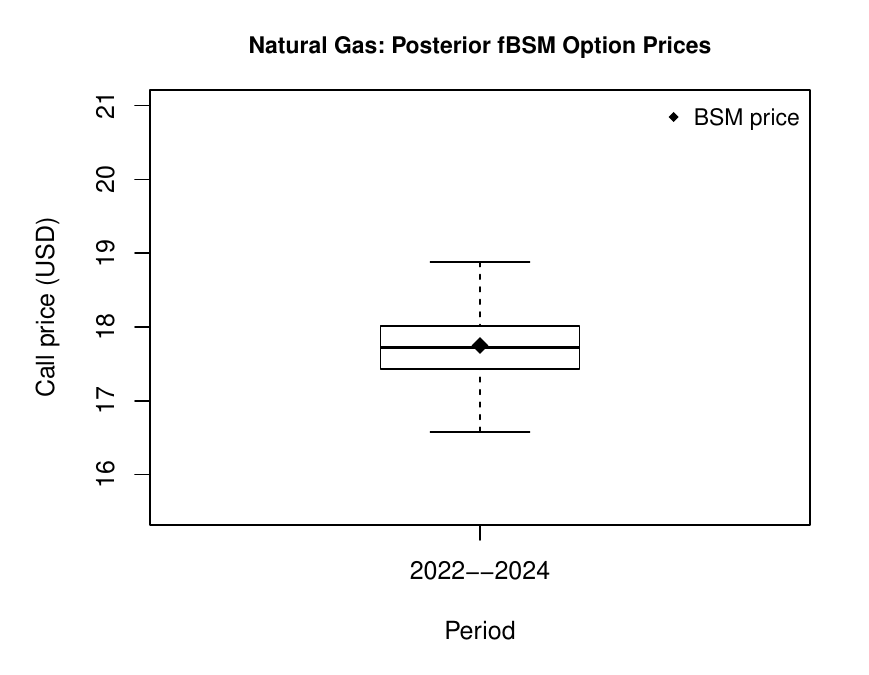}
\caption{Posterior distribution of option prices under the fractional
Black--Scholes model for natural gas. The Black--Scholes price is shown
for comparison.}
\label{fig:ng_boxplot}
\end{figure}

Figure~\ref{fig:ng_boxplot} illustrates the posterior distribution of
option prices. The posterior distribution is relatively wide,
reflecting substantial uncertainty in pricing outcomes. Although the classical Black--Scholes price lies within the posterior
pricing interval, it represents only a single point estimate and therefore
does not capture the full uncertainty associated with the joint estimation
of $(H,\sigma)$.

Overall, these results indicate that, in highly volatile markets, parameter
uncertainty can translate into meaningful variation in option values, even
in the absence of strong long-range dependence.

\section{Discussion and Conclusion}
\label{sec:conclusion}

This paper develops a Bayesian framework for estimating the Hurst parameter
and volatility in the fractional Black--Scholes model and examines their
impact on option pricing in energy markets. The empirical results provide
several important insights.

First, across all datasets considered, the estimated Hurst parameter
remains close to 0.5, suggesting weak evidence of long-range dependence
in returns despite elevated market volatility. This finding is consistent
across different assets and market conditions, indicating that persistence
plays a limited role in describing return dynamics. Although the estimated Hurst parameter for returns remains close to 0.5,
financial markets may still exhibit persistence in volatility dynamics,
which is consistent with the broader literature on volatility clustering
and rough volatility models. This suggests that fractional dynamics may
remain useful for uncertainty quantification even when strong long-range
dependence is not directly observed in return series.

However, the posterior distributions also demonstrate non-negligible
uncertainty surrounding these estimates. Consequently, the analysis does
not support imposing the classical Brownian motion assumption $H=0.5$
with certainty. From a risk management perspective, it is preferable to
allow for the possibility that the true persistence structure may deviate
from standard Brownian motion, even if the posterior center lies near
0.5. The Bayesian framework, therefore, provides a natural mechanism for
incorporating parameter uncertainty directly into inference and option
pricing, thereby avoiding overconfidence in a single fixed persistence
assumption. In this sense, the fractional model serves not only as a tool
for detecting strong long-range dependence, but also as a framework for
quantifying uncertainty about persistence in financial markets.

An important contribution of the proposed framework is not necessarily the detection of strong long-range dependence itself, but rather the quantification of uncertainty surrounding persistence. Although the posterior distributions of the Hurst parameter were centered near 0.5 in the empirical applications, the results do not support treating the classical Brownian motion assumption H = 0.5 as known with certainty. Instead, the Bayesian framework provides a probabilistic assessment of persistence by propagating uncertainty in the Hurst parameter directly into option pricing inference. Consequently, the methodology remains informative even when strong long-range dependence is not conclusively supported by the data, since uncertainty regarding persistence can still influence financial decision-making and risk assessment.

Second, volatility emerges as the primary driver of market behavior. In the
case of WTI crude oil, the comparison of two distinct periods reveals that
differences in market regimes are characterized by substantial changes in
volatility, while the Hurst parameter remains relatively stable. Similarly,
natural gas exhibits consistently high volatility, leading to greater
uncertainty in parameter estimates.

Third, these differences in volatility translate directly into differences
in option pricing uncertainty. While the median fractional Black--Scholes
prices remain close to the classical Black--Scholes values, the posterior
distributions reveal that the range of possible option prices can vary
considerably, particularly in high-volatility environments. This highlights
that classical pricing models may underestimate the uncertainty associated
with option valuation.

From a methodological perspective, the results emphasize the importance of
jointly modeling dependence and volatility. The Bayesian approach provides
a flexible framework that captures parameter uncertainty and avoids
over-reliance on point estimates. These findings are consistent with recent studies emphasizing the advantages
of Bayesian approaches for uncertainty quantification in long-memory
estimation problems \citep{MangalamLikens2025,Mangalam2025}. In contrast, methods based solely on moment conditions may be more sensitive to
volatility-driven variation, potentially leading to misleading
interpretations of persistence.

Overall, the findings suggest that, in financial markets, uncertainty in
volatility plays a more significant role than long-range dependence in
determining option prices. This has important implications for risk
management, as it indicates that accounting for parameter uncertainty is essential for accurate pricing and hedging decisions.

This study has several limitations. The return series are modeled using
a Gaussian fractional Gaussian noise likelihood with a constant volatility
parameter within each sample period. Consequently, the framework does not
explicitly account for stochastic volatility, jumps, heavy tails, or
structural breaks that may arise in financial markets. In addition, the
empirical analysis is based on daily data, which may mask microstructure
effects and intraday dependence patterns.

Future research may extend this framework to incorporate stochastic
volatility models, alternative prior structures, and heavy-tailed
likelihoods. Additional directions include applications to high-frequency
financial data and the integration of fractional dynamics with more
complex market mechanisms such as jumps, leverage effects, and
regime-switching behavior. Overall, the proposed Bayesian framework demonstrates that uncertainty quantification in fractional financial models can provide valuable insight into the stability and reliability of option pricing decisions under complex market conditions.

\end{document}